\documentclass[manuscript,screen,nonacm]{acmart}
\AtBeginDocument{%
  }

\begin{document}

\title{The Illusion of Competence: Self-Perceived Digital Literacy and AI Readiness Among European Secondary Students}

\author{Nicolas Rodriguez-Alvarez}
\orcid{0009-0002-6804-386X}
\affiliation{%
  \institution{Instituto de Educación Secundaria Parquesol}
  \country{Spain}
}

\author{Alan Martin Blanch-Marsolini}
\affiliation{%
  \institution{Instituto de Educación Secundaria Parquesol}
  \country{Spain}
}

\author{Samuel Vara-Gutierrez}
\affiliation{%
  \institution{Moate Community School}
  \country{Ireland}
}

\author{Hugo Gil-Garcia}
\affiliation{%
  \institution{Instituto de Educación Secundaria José Jiménez Lozano}
  \country{Spain}
}

\author{Javier Calzon-Dueñas}
\affiliation{%
  \institution{Instituto de Educación Secundaria Parquesol}
  \country{Spain}
}

\author{Fernando Rodriguez-Merino}
\orcid{0000-0002-3991-2563}
\authornote{Corresponding author}
\email{fernando.rodriguez.merino@uva.es}
\affiliation{%
  \institution{Universidad de Valladolid (UVA)}
  \city{Valladolid}
  \country{Spain}
}

\begin{abstract}
The ubiquitous presence of digital devices has cemented the 'Digital Native' paradigm, assuming inherent technological proficiency among contemporary youth. This multicenter study ($N=243$ European secondary students) challenges this narrative by investigating the gap between self-perceived digital literacy and actual technical readiness, including Artificial Intelligence (AI) interaction. Our findings reveal a severe Confidence-Competence Divide characterized by a collective Dunning-Kruger effect: students report near-maximum self-efficacy in passive digital consumption but exhibit a sharp decline when evaluating active technological creation and algorithmic logic. Crucially, an intra-pathway analysis demonstrates that the technological gender gap is not universal; rather, it emerges significantly exclusively within Technology-oriented classrooms ($p = 0.046$), indicating the persistence of 'stereotype threat' in formal STEM environments. Additionally, the study uncovers an 'AI Paradox' wherein students significantly overestimate their critical awareness of deepfakes and algorithmic biases compared to their operational AI skills, fostering a false sense of invulnerability against modern misinformation. Ultimately, supported by an overwhelming student demand ($76.5\%$) for pedagogical reform, this research concludes that dismantling this illusion of competence requires abandoning passive theoretical instruction in favor of hands-on, active technological creation.
\end{abstract}




\maketitle

\section{Introduction}
The term 'Digital Native' has permeated educational discourse for decades, creating a pervasive assumption that individuals born into the era of ubiquitous digital technology are inherently proficient in its use. However, current research suggests that this perceived fluency is often superficial. As argued by \citet{KIRSCHNER2017135}, familiarity with touchscreens and social media interfaces does not translate into deep technological literacy or algorithmic thinking. Furthermore, empirical evidence suggests a more nuanced reality beyond the popular narrative of tech-savvy youth. \citet{j.1475-682X.2009.00317.x} identified a 'second-level digital divide', showing that internet skills are not randomly distributed but are deeply influenced by gender and socioeconomic factors, which leaves many students limited to basic, non-professional uses. The unprecedented integration of Artificial Intelligence into the educational landscape is currently exposing this significant competency gap. This phenomenon prompts a critical re-evaluation of whether the 'digital native' paradigm accurately reflects students' ability to master complex technological tools in a rapidly automating world. Building upon this theoretical framework, our research explores the 'Confidence-Competence Divide' among 243 European secondary students. We examine how the aforementioned differences manifest in a multicultural sample, specifically looking at how academic pathways and gender influence the perceived readiness for advanced digital tasks such as programming and CAD. Ultimately, this study proposes that the current generation may be suffering from a collective Dunning-Kruger effect: overestimating their digital skills based on their proficiency as passive consumers, while remaining functionally unequipped for active technological creation and logical troubleshooting.

\section{Research Questions}
The primary goal of this research is to address the following questions regarding digital competence and AI readiness:
\begin{enumerate}
    \item \textbf{RQ1:} How does students' self-perceived digital competence vary between passive technological consumption and active algorithmic creation?
    \item \textbf{RQ2:} To what extent do academic pathways (e.g., STEM vs. Arts) and gender influence the manifestation of the technological confidence gap?
    \item \textbf{RQ3:} Is there a disconnect between students' operational AI usage and their critical awareness of algorithmic biases and deepfakes?
    \item \textbf{RQ4:} What pedagogical reforms do secondary students identify as necessary to improve formal IT education?
\end{enumerate}

\section{Methodology}
\subsection*{Study Design and Participants}
A cross-sectional, multicenter study was conducted to evaluate the digital and algorithmic self-efficacy of secondary school students. The initial sample consisted of students aged 12 to 16 from three educational institutions located in Spain (Valladolid and Zamora) and Ireland (County Westmeath). After rigorous data cleaning procedures, which included the identification and removal of duplicate entries, the final validated dataset comprised $N = 243$ participants. The sample encompassed diverse academic pathways, allowing for a comparative analysis between students enrolled in Technology, Sciences, and Arts and Humanities electives.

\subsection*{Instrument Design}
Data was collected using a structured, self-administered questionnaire. The core of the instrument consisted of a 5-point Likert scale (ranging from 1: 'Strongly Disagree' to 5: 'Strongly Agree'), designed to assess self-perceived competence across multiple domains. These dimensions included: (1) passive digital consumption and basic interface navigation, (2) digital privacy and security, (3) logical troubleshooting and algorithmic thinking, (4) active technological creation (e.g., programming and CAD), (5) Artificial Intelligence (AI) readiness, and (6) student evaluation of the current IT curriculum.

\subsection*{Statistical Analysis}
The anonymized data was processed and analyzed using Python (Pandas and SciPy libraries). Descriptive statistics (means and standard deviations) were calculated to establish the baseline self-efficacy in each technological dimension. To test the research hypotheses, inferential statistical analyses were performed. Independent samples t-tests, applying Welch's correction to account for unequal variances, were utilized to evaluate gender and academic pathway disparities. Additionally, paired-samples t-tests were conducted to compare intra-subject variances, specifically contrasting operational AI usage with critical AI awareness. The threshold for statistical significance was set at $\alpha = 0.05$.

\subsection*{Data Collection Procedure}
To ensure methodological rigor and mitigate potential peer-coercion or social desirability bias, the student-authors were completely detached from the survey administration process. Data collection was officially coordinated through the administrative headship (\textit{Jefatura de Estudios}) of the participating institutions. The survey was administered in-person by the students' respective homeroom teachers during dedicated tutorial sessions. This double-blind approach guaranteed that participation was entirely voluntary, strictly anonymous, and free from any peer-to-peer influence.

\subsection*{Sex and Gender Reporting}
In accordance with the Sex and Gender Equity in Research (SAGER) guidelines, this study relies on self-reported gender identity as a social construct rather than assigned biological sex. The original survey instrument was designed with inclusive language, offering options beyond the binary categorization (e.g., 'non-binary', 'other', or 'prefer not to say'). However, the number of respondents who selected these non-binary options was statistically negligible ($n = 2$ out of $N = 243$, representing $< 1\%$ of the sample). To ensure strict compliance with the General Data Protection Regulation (GDPR) and the Spanish LOPDGDD, and to completely eliminate any risk of de-anonymization for these minority participants within specific schools, these two data points were excluded from the comparative demographic analysis. Consequently, the gender-based variance reported in this study (independent t-tests) focuses exclusively on the robust cohorts of self-identified male and female students ($n = 241$).

\section{Results}

\begin{figure*}[t]
    \centering
    \includegraphics[width=\textwidth]{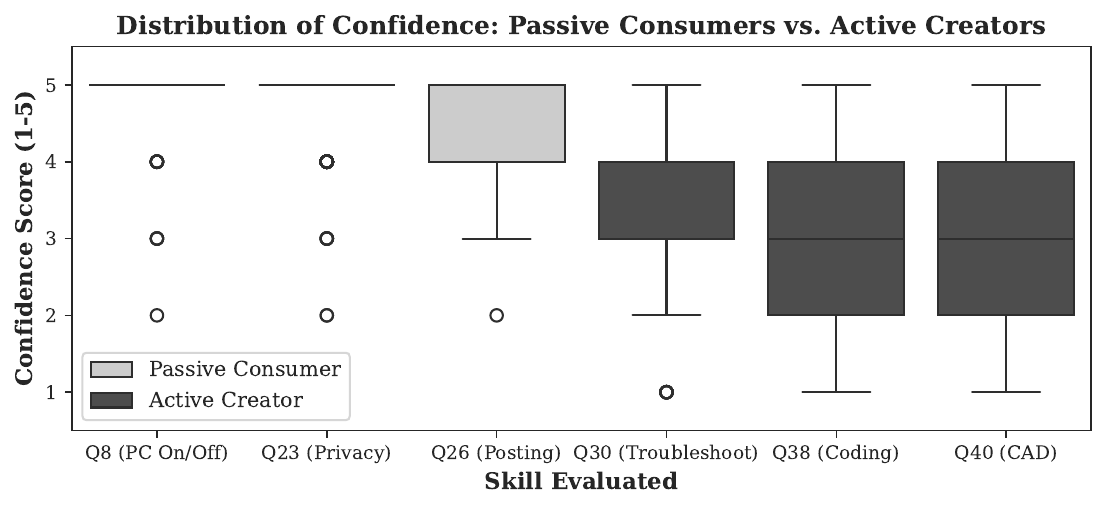}
    \caption{Distribution of self-perceived confidence: Passive Consumers vs. Active Creators. The compressed upper interquartile range for basic tasks contrasts heavily with the high variance and lower medians of technical skills.}
    \Description{A collection of side-by-side box plots illustrating the distribution of confidence scores (1-5) for six skills, grouped into Passive Consumers (light gray) and Active Creators (dark gray). Skills such as turning a PC on/off and privacy show a very high, narrow distribution with medians at 5 and few outliers. Conversely, active creation skills like coding, CAD, and troubleshooting show much lower and more dispersed confidence, with medians around 3-4, wider interquartile ranges, and lower outliers down to 1, showing a marked contrast between these two skill types.}
    \label{fig:skills}
\end{figure*}

A total of 243 students from Spain and Ireland participated in the survey. The data collection revealed a striking disparity in self-perception depending on the complexity of the task. For basic functional skills, such as turning on a computer (Q8) or managing privacy (Q23), students reported near-maximum confidence levels ($M=4.88$ and $M=4.81$, respectively). Nevertheless, when evaluating active creation skills---such as using CAD software ($M = 2.86$), programming ($M = 2.96$), and understanding logical troubleshooting steps ($M = 3.29$)---the self-reported scores dropped significantly. This sharp decline in confidence, illustrated in Figure \ref{fig:skills}, highlights a severe perceived inability to engage with algorithmic logic. While low self-efficacy does not inherently equate to absolute incompetence, it exposes a critical barrier in algorithmic literacy: students feel fundamentally unequipped to transition from passive consumers to active problem-solvers.

\begin{figure}[htbp]
    \centering
    \includegraphics[width=\columnwidth]{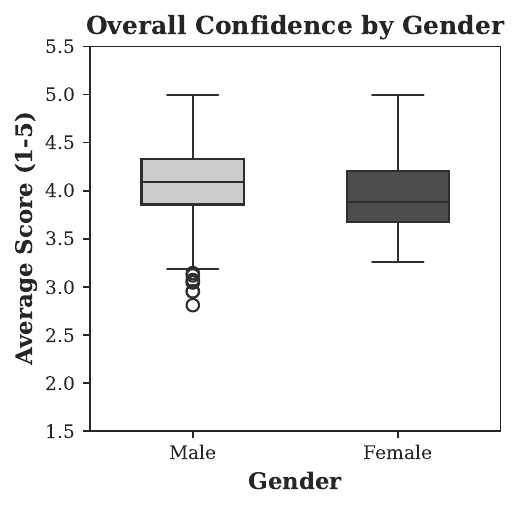}
    \caption{Overall digital confidence distribution by gender. The upward shift in the male interquartile range indicates a systemic overestimation of technical readiness compared to female students.}
    \Description{Two side-by-side box plots, in light gray for Male and dark gray for Female, displaying overall confidence scores on a 1-5 scale. The distribution for Males shows a slightly higher median score of approximately 4.1 compared to Females at approximately 3.9. While both genders share similar upper quartile and maximum scores, the Male distribution is more dispersed with visible lower outliers down to approximately 2.8, indicating greater variability in confidence among male students.}
    \label{fig:gender}
\end{figure}
    
Furthermore, demographic analysis revealed a systemic gender gap in overall technological self-efficacy. Despite sharing identical educational environments, male students consistently reported higher generalized digital confidence ($M = 4.05$) compared to their female counterparts ($M = 3.89$). This difference was found to be statistically significant ($p = 0.001$). As depicted in Figure \ref{fig:gender}, this disparity is not driven by isolated statistical outliers, but rather by an upward shift of the entire interquartile range for male respondents. This structural displacement suggests that female adolescents preemptively underestimate their broad technical readiness, a psychological barrier that aligns with established literature on the early onset of the STEM gender divide \citep{10.1177/0956797617741719,Wang2017}.

Finally, cross-tabulation with students' chosen academic electives (Q6) highlighted the significant influence of the curriculum on specific digital competencies. Students enrolled in Technology-oriented pathways reported the highest self-efficacy in active creation skills ($M = 3.43$). In stark contrast, students in Arts and Humanities pathways exhibited the lowest confidence in these logical tasks ($M = 3.04$). An independent t-test confirmed that this variance in technical readiness between Technology and Arts students is highly statistically significant ($p < 0.001$). This underlines that mere exposure to digital devices is insufficient; formal, specialized instruction is critical to bridging the confidence-competence divide.

\begin{figure*}[htbp]
    \centering
    \includegraphics[width=\textwidth]{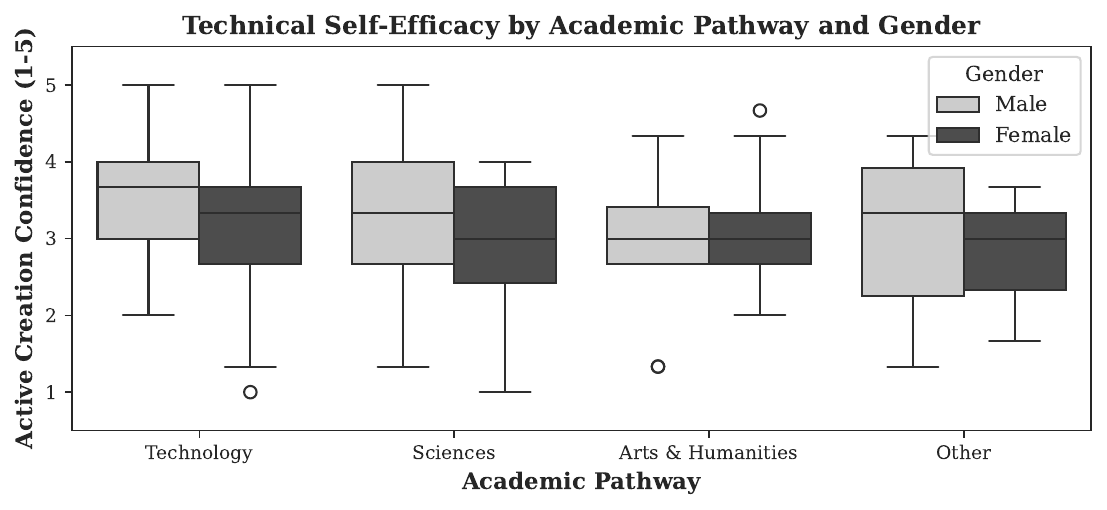}
    \caption{Technical Self-Efficacy by Academic Pathway and Gender. The statistical parity found in Arts and Sciences contrasts with the significant gender gap emerging exclusively within the Technology pathway.}
    \Description{A set of grouped box plots titled 'Technical Self-Efficacy by Academic Pathway and Gender', showing active creation confidence on a 1-5 scale. Confidence is broken down by four academic pathways: Technology, Sciences, Arts & Humanities, and Other, with a color legend for gender: Male (light gray) and Female (dark gray). Across all pathways except Arts & Humanities, Males show slightly higher median confidence than Females. In Arts & Humanities, the medians are very similar, around 3, with Female having a high outlier and Male having a lower outlier. Generally, all scores are in the 2-4 range, visibly lower than overall confidence scores.}
    \label{fig:pathway}
\end{figure*}

To further investigate the nuances of the technological gender gap, an intra-pathway analysis was conducted. Independent t-tests (Welch's correction) were applied to evaluate male-female self-efficacy differences strictly within their chosen academic fields. The results demonstrated that the gender gap is not uniformly distributed across the student body. In the Arts and Humanities pathway ($n=48$), no significant gender difference in active creation confidence was found ($p = 0.841$). Similarly, the Sciences pathway ($n=75$) exhibited statistical parity between genders ($p = 0.142$). However, within the Technology-oriented pathway ($n=111$), a statistically significant gender gap emerged ($p = 0.046$), as illustrated in Figure \ref{fig:pathway}, with male students reporting disproportionately higher self-efficacy than their female classmates.

\begin{figure}[htbp]
    \centering
    \includegraphics[width=\columnwidth]{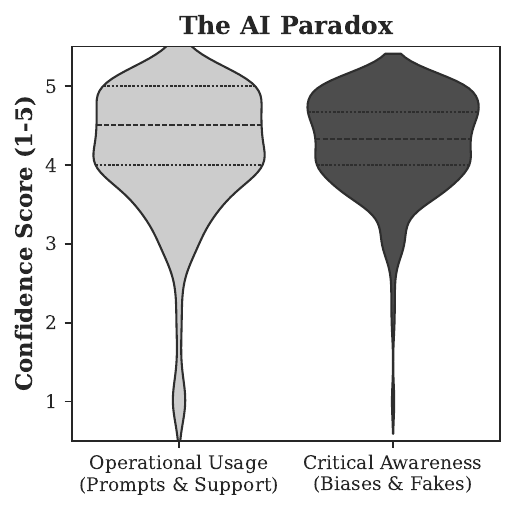}
    \caption{The AI Paradox. Students report higher confidence in their critical awareness (detecting biases and deepfakes) than in their operational usage of AI tools.}
    \Description{A two-sample violin plot with internal dashed quartile and median lines, in light gray for 'Operational Usage (Prompts & Support)' and dark gray for 'Critical Awareness (Biases & Fakes)', comparing confidence scores on a 1-5 scale. The operational violin is visibly wider at the top, reflecting a high and narrow concentration of scores around 4-5, with a median of approximately 4.5 and a long lower tail. The critical awareness violin is narrower at the top and broader in the middle, indicating scores are more dispersed across the middle range, with a median of approximately 4.3 and a more symmetrical shape, highlighting a clear confidence gap between simple operational use and more advanced critical skills.}
    \label{fig:ai}
\end{figure}

In addition to traditional digital literacy, the study evaluated students' self-perceived Artificial Intelligence (AI) readiness. Paired-samples t-tests were conducted to compare operational AI usage (e.g., writing effective prompts) against critical AI awareness (e.g., understanding biases and identifying deepfakes). Interestingly, the results revealed a statistically significant difference ($p = 0.015$), with students reporting higher confidence in their critical evaluation skills ($M = 4.28$) than in their operational skills ($M = 4.15$). This exceptionally high self-perception of critical awareness contrasts with the complex reality of modern AI-generated misinformation (see Figure \ref{fig:ai}).

\begin{figure*}[htbp]
    \centering
    \includegraphics[width=\textwidth]{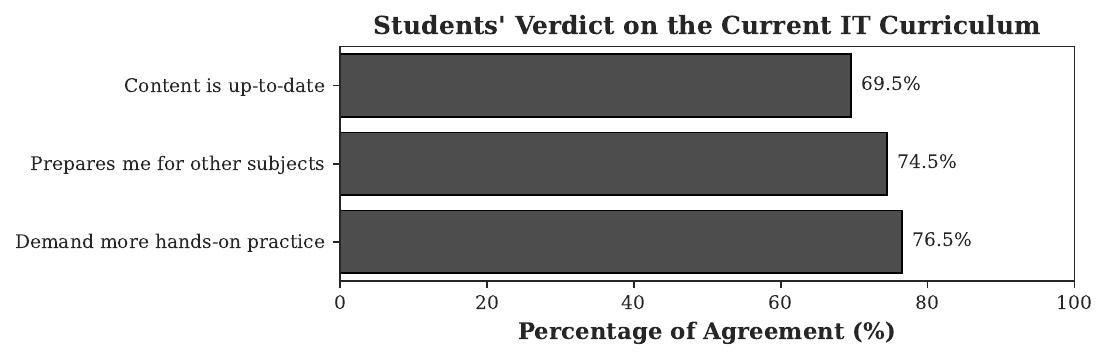}
    \caption{Students' verdict on the current IT curriculum, highlighting a strong demand for hands-on pedagogical approaches.}
    \Description{A horizontal bar chart titled 'Students' Verdict on the Current IT Curriculum', showing a very high percentage of agreement across three statements. 'Content is up-to-date' has ~69.5\% agreement, 'Prepares me for other subjects' ~74.5\%, and 'Demand more hands-on practice' is the highest at ~76.5\%, indicating a widespread desire for increased practical learning.}
    \label{fig:curriculum}
\end{figure*}

Finally, regarding the evaluation of their formal technological education, the student body conveyed a clear mandate for pedagogical reform. While a substantial majority ($74.5\%$) felt that their computer science classes adequately prepared them for the digital requirements of other subjects, an overwhelming $76.5\%$ of respondents explicitly demanded a shift towards more hands-on, practical computer activities rather than theoretical instruction, as summarized in Figure \ref{fig:curriculum}.

\section{Discussion}
The findings of this multicenter study challenge the pervasive paradigm of the 'Digital Native', providing empirical evidence that ubiquitous access to technology does not inherently cultivate deep digital literacy. Our data reveals a severe Confidence-Competence Divide characterized by a collective Dunning-Kruger effect: students report near-maximum self-efficacy in passive consumption and basic interface navigation (e.g., social media privacy, basic OS operations) but exhibit a sharp decline in confidence when confronted with active technological creation and algorithmic logic. This supports the assertions of \citet{KIRSCHNER2017135}, confirming that the superficial fluency derived from touchscreen familiarity often masks a profound inability to transition from digital consumers to active problem-solvers.

Furthermore, the intra-pathway analysis reveals a concerning paradox regarding the technological gender gap. Strikingly, the disparity in active creation confidence is absent in the Arts and Humanities pathway but becomes highly significant within the Technology pathway ($p = 0.046$). This phenomenon strongly aligns with the concept of \textit{stereotype threat} in computer science education. As established by \citet{2016-48466-001}, the structural 'masculine culture' often associated with STEM environments can diminish female students' sense of belonging. Even when female students actively choose technological pathways and receive identical formal instruction, pervasive cultural stereotypes regarding innate male technical brilliance can severely undermine their confidence \citep{Beyer03072014}. Consequently, merely enrolling girls in STEM classes is insufficient; the educational environment itself must actively dismantle these implicit biases to prevent the amplification of the imposter syndrome among female adolescents.

This illusion of competence also extends to the realm of Artificial Intelligence. Our analysis of 'AI Readiness' exposes a critical paradox: students report significantly higher confidence in their critical awareness, such as detecting deepfakes and AI biases ($M = 4.28$), than in their operational usage of AI tools ($M = 4.15$). This exceptionally high self-perception of critical evaluation skills contrasts with the complex reality of modern AI-generated misinformation. Recent psychological studies suggest that humans systematically overestimate their ability to differentiate authentic content from sophisticated audiovisual deepfakes. The familiarity of this generation with basic social media filters appears to create a false sense of invulnerability, leading to an 'illusion of competence' where young users severely underestimate the sophistication of modern algorithmic manipulation \citep{hashmi2024unmaskingillusionsunderstandinghuman}.

Ultimately, these findings translate into a clear mandate for pedagogical reform, echoed by the students themselves. While respondents value their IT education, an overwhelming 76.5\% demand a paradigm shift from theoretical instruction toward hands-on, practical computer activities. To dismantle the Dunning-Kruger effect in both traditional computing and AI interaction, educational institutions must abandon passive instructional models. Fostering true digital resilience requires curricula centered on algorithmic trial-and-error, logical troubleshooting, and active technological creation.

\section{Conclusions}
This study empirically challenges the assumption that the 'Digital Native' generation possesses inherent technological fluency. By analyzing the self-perceived digital competence of 243 European secondary students, we identified a profound Confidence-Competence Divide. Students exhibit near-maximum confidence in basic, passive digital tasks but report severe insecurity when faced with active creation and algorithmic logic, evidencing a widespread Dunning-Kruger effect.

Crucially, our findings deconstruct the technological gender gap, revealing it is not a uniform generational trait but a highly contextual phenomenon. The statistical parity observed in Arts and Sciences contrasts sharply with the significant confidence gap found specifically within the Technology pathway ($p = 0.046$), highlighting the persistent impact of stereotype threat in STEM environments. Furthermore, the 'AI Paradox' identified in this study---where students overestimate their critical awareness of deepfakes and biases compared to their operational skills---warns of a new dimension of digital vulnerability.

Ultimately, this research serves as a call to action for educational policymakers. With 76.5\% of students explicitly demanding more hands-on practice, it is imperative to shift the IT curriculum from passive software navigation toward active computational thinking and algorithmic problem-solving. Only through specialized, practical instruction can we empower the current generation to transition from mere consumers of technology to resilient, critical creators.

\section{Limitations}
The scope of this study is limited to a convenience sample of students from one secondary school in Valladolid (Spain), one in Zamora (Spain), and one in County Westmeath (Ireland). While this multicenter approach provides valuable diverse insights, the results should be interpreted cautiously. Future research should aim to replicate this methodology across a broader, randomized demographic of educational institutions and regions to further confirm the generalizability of these findings.

\section*{Ethical Statement}
All participants provided informed consent prior to data collection in accordance with the GDPR and LOPDGDD. As the survey was completely anonymous, non-interventional, and collected no sensitive personal data, the study was exempt from formal Institutional Review Board (IRB) approval.

\section*{Funding}
This research did not receive any specific grant from funding agencies in the public, commercial, or not-for-profit sectors.

\section*{Data Availability Statement}
In alignment with Open Science principles and to ensure the reproducibility of this research, the fully anonymized dataset ($n=243$) utilized in this study has been released into the public domain under a CC-BY 4.0 license. The dataset, encompassing all demographic variables and Likert-scale responses, along with its comprehensive codebook, is freely accessible via the Zenodo repository \citep{rodriguez_alvarez_2026_18822871}.

\begin{acks}
\vspace{1em}
\noindent \textbf{Declaration of Generative AI and AI-assisted technologies in the manuscript preparation process:} 
During the preparation of this work, the authors used Google Gemini in order to improve the English academic phrasing and ensure optimal readability of the manuscript. After using this tool, the authors reviewed and edited the content as needed and take full responsibility for the content of the published article.
\end{acks}

\bibliographystyle{ACM-Reference-Format}
\bibliography{cas-refs}

\end{document}